# Operando multidimensional spectroscopy reveals A-site-dependent carrier cooling in perovskite solar cells

Edoardo Amarotti[1,2], Luca Bolzonello[2], Qi Shi[1], Sun-Ho Lee[3], Torbjörn Pascher[1], Donatas Zigmantas[1], Niek van Hulst[2,5], Nam-Gyu Park[3,4], Tönu Pullerits[1,*]

[1]Department of Chemical Physics and NanoLund, Lund University, P.O. Box 124, Lund, 22100, Sweden.

[2]ICFO - Institut de Ciències Fotòniques, The Barcelona Institute of Science and Technology, Castelldefels, 08860, Spain.

[3]School of Chemical Engineering, Sungkyunkwan University, Suwon, 16419, Republic of Korea.

[4]SKKU National Lab for Intelligent Energy Solution Technology (SIEST), Sungkyunkwan University, Suwon, 16419, Republic of Korea.

[5]ICREA—Institució Catalana de Recerca i Estudis Avançats, The Barcelona Institute of Science and Technology, Castelldefels, 08010, Spain.

[*]Corresponding author. E-mail: tonu.pullerits@chemphys.lu.se

## Abstract

Understanding how photogenerated carriers dissipate excess energy in operating perovskite solar cells is essential for connecting ultrafast photophysics with photovoltaic function and provides design principles for engineering next-generation cell architectures. Here we use photocurrent-detected two-dimensional electronic spectroscopy (PC-2DES) to resolve energy-dependent carrier relaxation in fully encapsulated, functioning metal halide perovskite solar cells. By comparing devices with identical architecture but different absorber compositions — $MAPbI_3$, mixed FAMA, and $FAPbI_3$ — we directly follow the redistribution of photoexcited carriers from initially populated high-energy states toward lower-energy band-edge states. The multidimensional photocurrent response reveals a cascade-like intraband cooling process whose rate depends strongly on absorber composition, with the slowest relaxation in MA-based devices, intermediate behaviour in mixed-cation devices, and fastest relaxation in FA-based devices. A reduced kinetic model incorporating phonon-mediated intraband scattering, supplemented by a phenomenological many-body contribution, captures the main energy-dependent trends. These results establish action-detected multidimensional spectroscopy as a device-level probe of ultrafast energy dissipation and show that subtle changes in perovskite composition can substantially reshape the carrier relaxation pathways that precede charge extraction.

## Introduction

Metal halide perovskite solar cells combine strong absorption,[1,2] long carrier lifetimes,[3,4] and efficient charge extraction,[5,6] but the microscopic processes that connect photoexcitation to device output remain incompletely understood under operating conditions.[7] Immediately after above-bandgap excitation, carriers possess excess kinetic energy that must be dissipated through carrier–carrier and carrier–phonon interactions before charge extraction and recombination are complete.[8–11] This intraband cooling controls how electronic energy is transferred to the lattice and interfaces, and may therefore influence non-radiative losses, local heating, ion motion, and operational stability.[7,12]

The A-site cation provides a particularly sensitive handle on these relaxation pathways.[13,14] Although the band-edge states of lead halide perovskites are primarily derived from the inorganic Pb–halide framework, the organic cation modifies lattice dimensions, octahedral distortions, dielectric screening, dynamic disorder, and vibrational coupling.[15–17] Replacing methylammonium (MA) with formamidinium (FA), or using mixed A-site compositions, can therefore alter not only the bandgap and phase stability[18] but also the phonon-mediated pathways through which hot carriers dissipate excess energy.[19]

Ultrafast optical spectroscopy has provided extensive information on carrier thermalization,[9] hot-phonon effects,[20] intraband relaxation,[21] and charge separation[22] in perovskite films and crystals. However, measurements on simplified samples do not fully capture the environment of a complete photovoltaic device, where charge-transport layers, built-in fields, interfaces, electrodes, and encapsulation can all reshape carrier dynamics. Directly resolving energy-dependent relaxation in an operating device remains challenging because conventional optical detection is often compromised by multilayer interference, scattering, weak differential signals, and the difficulty of connecting optical observables to extracted charge. This is the central experimental gap addressed here.

Photocurrent-detected two-dimensional electronic spectroscopy (PC-2DES) offers a route around these limitations.[23,24] By correlating excitation and detection energies while measuring the generated photocurrent, PC-2DES follows the evolution of photoexcited populations through an electrical observable.[25] The method therefore combines the spectral and temporal resolution of coherent multidimensional spectroscopy with direct sensitivity to the states and relaxation pathways that contribute to charge extraction in a working solar cell.

Here we apply operando PC-2DES to fully encapsulated perovskite solar cells with identical device architecture but different absorber compositions: $FAPbI_3$, mixed FAMA, and $MAPbI_3$. The multidimensional photocurrent response reveals a cascade-like redistribution of carriers from initially excited high-energy states toward lower-energy band-edge states. Comparing the three compositions shows a clear composition-dependent hierarchy in the cooling dynamics, with the slowest relaxation in MA-based devices and the fastest in FA-based devices. A reduced kinetic model incorporating phonon-mediated intraband scattering and a phenomenological many-body contribution provides a physically motivated description of the energy-dependent dynamics. These results establish PC-2DES as a device-level probe of ultrafast energy dissipation and show that subtle changes in absorber compositions can substantially alter the early-time carrier relaxation pathways that precede charge extraction.

## Operando PC-2DES of perovskite solar cells with different absorber compositions

To examine how absorber composition influences ultrafast carrier relaxation in working devices, we compared fully encapsulated perovskite solar cells with identical n–i–p architecture and charge-transport layers, using $FAPbI_3$, mixed FAMA and $MAPbI_3$ absorbers. This design minimizes device-level variations and allows the measured photocurrent dynamics to be compared directly across the three absorber compositions.

Fig. 1a summarizes the experimental setup. The devices were measured under short-circuit conditions using PC-2DES. A sequence of four phase-modulated ultrashort laser pulses excites the solar cell, and the resulting photocurrent signal is presented as a function of excitation energy, detection energy and population time. Details are provided in Supplementary Note (SN) 1. In this action-detected geometry, the measured multidimensional response is directly weighted by the ability of photoexcited states to contribute to extracted charge, making the experiment sensitive to relaxation pathways that are relevant for device operation, see SN2.

The steady-state absorption spectra displayed in Fig. 1b confirm the expected compositional shift in optical response, with $MAPbI_3$ blue-shifted relative to the mixed-cation device and $FAPbI_3$. The excitation spectra used for the PC-2DES measurements were chosen to cover the near-band-edge absorption region of each device. Thus, while the absolute excitation windows differ slightly between compositions, the experiment probes the redistribution of carriers within the optically accessible conduction-band manifold in each working solar cell.

This combination of matched device architecture, composition-dependent absorption and photocurrent-detected multidimensional spectroscopy provides the basis for comparing carrier cooling across the three materials. We first examine the evolution of the two-dimensional photocurrent maps, which directly visualize the redistribution of spectral weight during the population time, and then quantify the energy-dependent relaxation dynamics through selected excitation–detection coordinates.

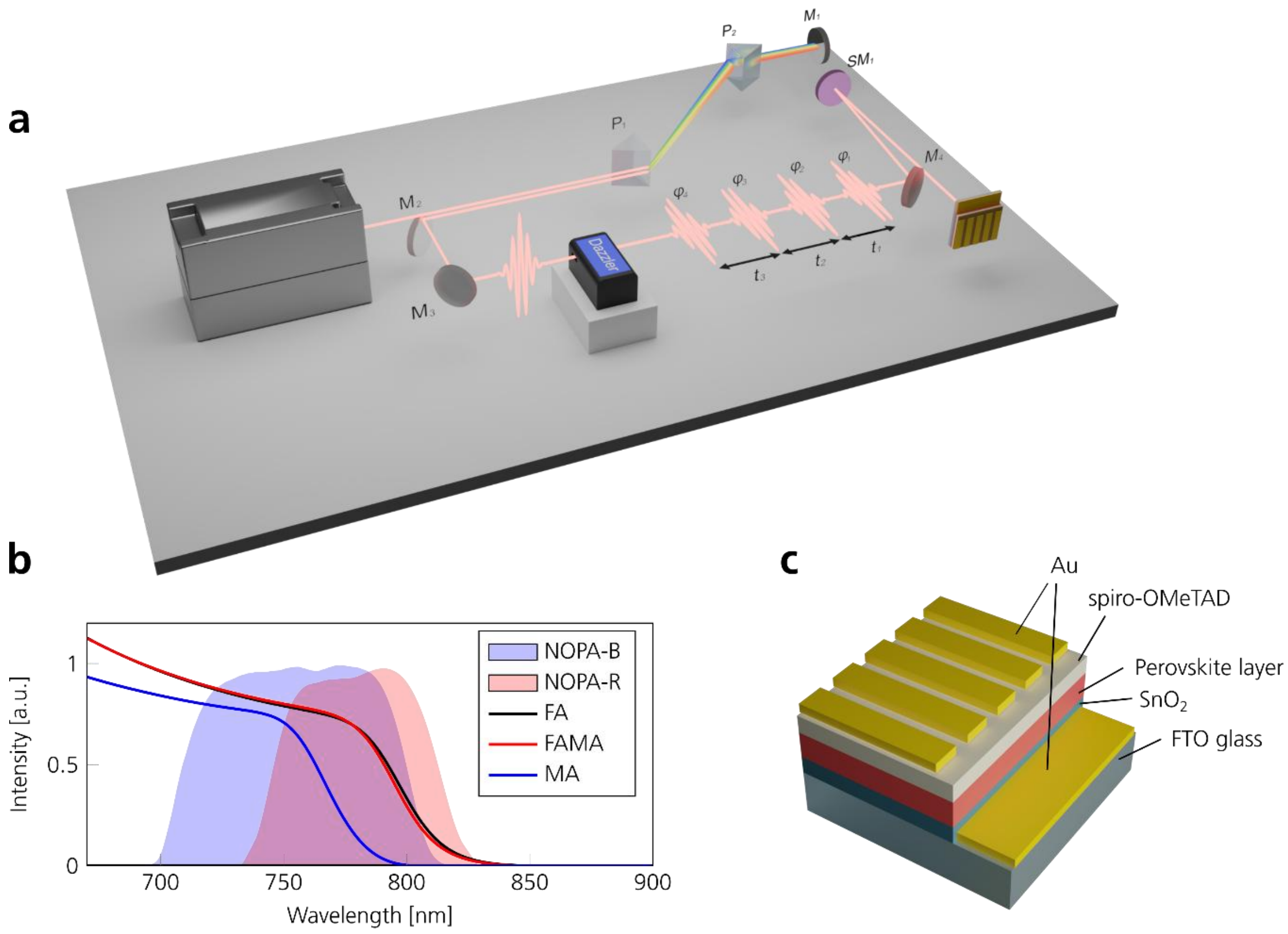

**Figure 1| Operando photocurrent-detected two-dimensional electronic spectroscopy of perovskite solar cells.**

**a**, Schematic of the PC-2DES experiment applied to a fully encapsulated perovskite solar cell. Pulses from noncollinear optical parametric amplifier (NOPA) are compressed to 20 fs using a combination of a fused silica prism compression, $P_1$ and $P_2$. An acousto-optic programmable dispersive filter (Dazzler) shapes a single pulse to four phase-modulated pulses with precise control of three time delays. A spherical mirror $SM_1$ focuses the pulses to the solar cell, and the resulting photocurrent is detected as the action signal. Fourier transform over $t_1$ and $t_3$ yields excitation and detection energy, while $t_2$ provides population time dependence.

**b**, Steady-state absorption spectra of FA-, FAMA-, and MA-based devices, showing a progressive blue shift from FA to MA compositions. The shaded areas represent the NOPA spectra employed: the NOPA-R, red, has been used to perform the experiment on the FA and FAMA devices, while the NOPA-B, blue, was used to study the MA device.
**c**, Device architecture used for all measurements, illustrating the n–i–p stack and encapsulation.

## Energy-resolved carrier relaxation revealed by PC-2DES

Fig. 2 presents representative PC-2DES maps recorded for the $FAPbI_3$, mixed FAMA and $MAPbI_3$ devices at early and later population times. At 50 fs, after pulse-overlap artefacts have largely decayed, all three compositions show dominant near-diagonal features, indicating that the photocurrent response initially reflects the population of states within the excitation bandwidth. The early-time maps therefore provide a reference for the initially prepared carrier distribution in each device.

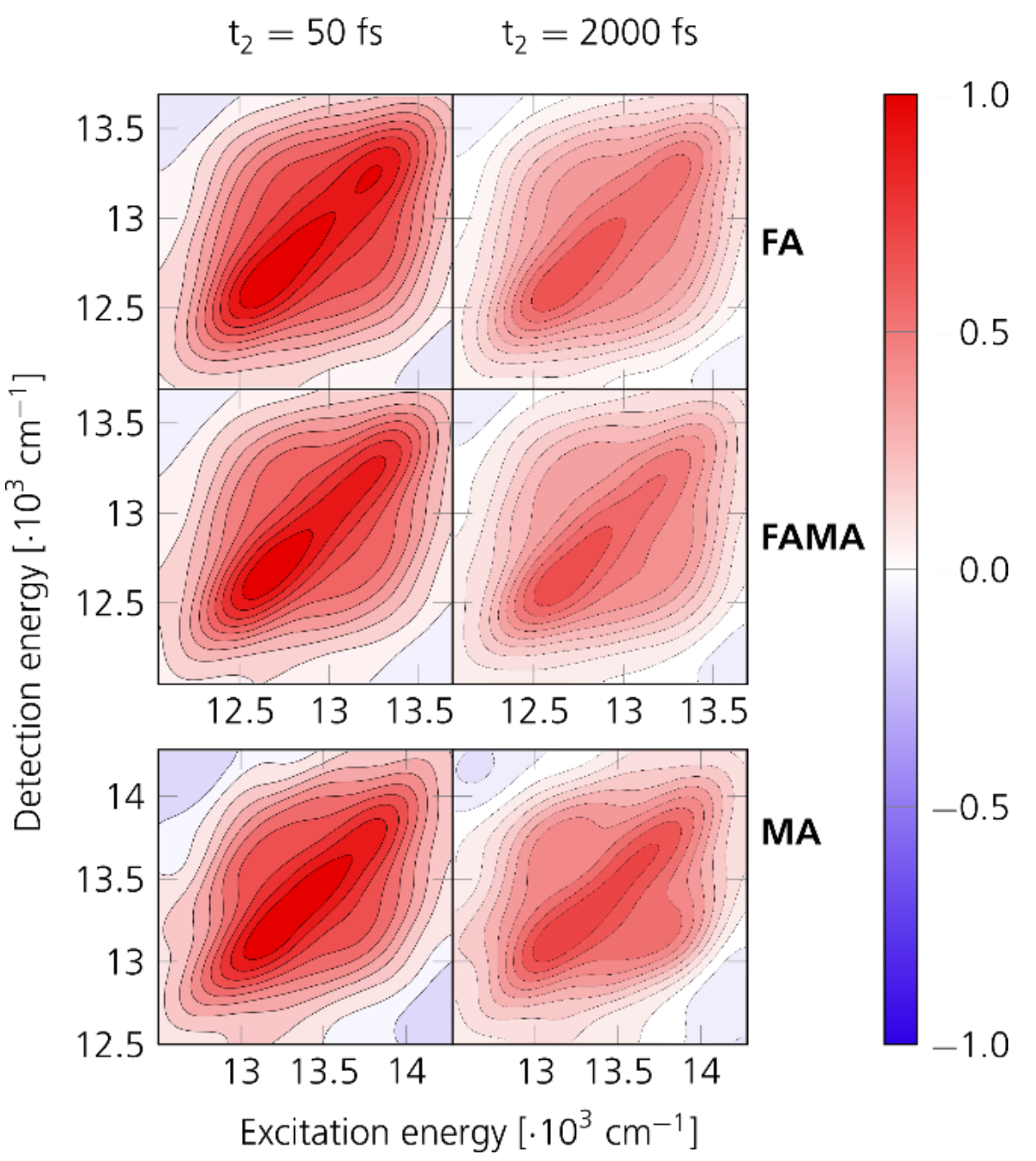

**Figure 2 | Two-dimensional photocurrent maps revealing carrier relaxation.**

Representative PC-2DES maps recorded for FA-, FAMA-, and MA-based perovskite solar cells at population times of 50 fs and 2 ps. At early times, all compositions exhibit dominant diagonal features corresponding to prompt population of conduction-band states. At later times, off-diagonal features emerge below the diagonal, indicating intraband relaxation toward lower-energy states. Differences in spectral evolution highlight the influence of absorber composition.

As the population time increases, the spectra evolve away from this initial diagonal response. At 2 ps, additional signal has appeared below the diagonal, corresponding to detection at lower energy than excitation. This below-diagonal response is the key experimental signature of energy redistribution: carriers initially prepared at higher energies contribute at later times to lower-energy photocurrent-detected states. The growth and spectral extent of this response differ between the three absorber compositions, indicating that the intraband relaxation pathways are modified by the perovskite lattice environment.

Because absolute PC-2DES amplitudes can be affected by differences in absorption strength, device photocurrent, excitation spectrum and extraction efficiency, the most robust comparison is the time evolution within each composition, rather than the absolute signal magnitude between different devices. We therefore use the 50 fs map as an early-time reference and compare how spectral weight redistributes during the population time. This analysis emphasizes the emergence of below-diagonal features associated with carrier cooling while reducing the influence of static device-to-device amplitude differences.

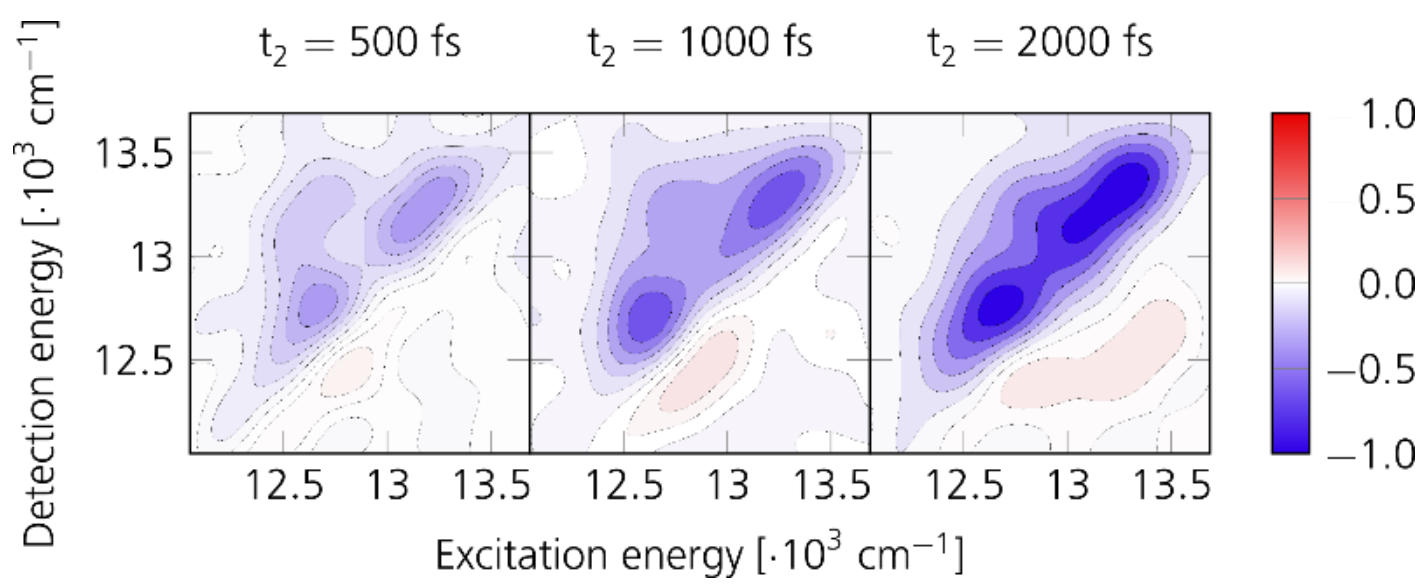

**Figure 3 | Differential PC-2DES maps highlighting relaxation dynamics**

Differential PC-2DES maps of FAMA obtained by subtracting the 50 fs reference spectrum from maps at longer population times. The subtraction enhances visibility of relaxation-driven population transfer. The evolution from diagonal-dominated features (blue) to horizontally extended signals (red) reveals redistribution of carriers toward the conduction-band minimum.

Fig. 3 shows differential maps obtained by subtracting the 50 fs spectrum from maps recorded at later population times. This differential analysis suppresses contributions that are already present at early delay, including static or slowly varying components that may arise from incoherent population-mixing in action-detected measurements.[26–28] For the details see SN 3. The resulting difference maps highlight the population-time-dependent redistribution of photocurrent-detected nonlinear response from the initially excited energy region toward lower detection energies. This below-diagonal growth provides a device-relevant measure of carrier redistribution, while the action-detected signal remains weighted by the nonlinear optical response and charge-extraction efficiency.

The differential maps reveal a common qualitative relaxation pathway across all three devices: an initially diagonal response gives way to below-diagonal and more horizontally extended features as carriers relax toward lower-energy states. However, the rate of this evolution is composition dependent. The FA-based device shows the most rapid development of low-energy response, the mixed-cation device displays intermediate behaviour, and the MA-based device evolves more slowly over the measured population time window. These trends show that differences in absorber composition are accompanied by changes not only in the optical bandgap but also in the timescale over which excess carrier energy is dissipated in working solar cells.

The 2D maps therefore provide a direct visual indication of composition-dependent intraband cooling. In the following section, we quantify this behaviour by extracting population time traces at selected excitation–detection coordinates, allowing the energy-dependent delay in low-energy photocurrent buildup to be compared across the three absorber compositions.

### Tracking intraband cooling across excitation and detection energies

To quantify the carrier redistribution visualized in the two-dimensional maps, we extracted population time traces from selected excitation–detection coordinates of the PC-2DES spectra. These coordinates were chosen to follow how carriers initially prepared at different excess energies contribute to photocurrent detected near lower-energy states. This analysis converts the spectral evolution of the maps into energy-resolved kinetic traces that can be compared directly across the three absorber compositions.

Fig. 4 shows representative traces obtained from the $FAPbI_3$, mixed FAMA and $MAPbI_3$ devices. For excitation close to the detection energy, the photocurrent response rises promptly, consistent with direct population of states that already contribute to the selected detection channel. By contrast, when the excitation energy is increased while the detection energy is kept near the lower-energy region, the signal onset is delayed. This delay reflects the finite time required for carriers prepared at higher energies to relax toward the states that dominate the low-energy photocurrent response.

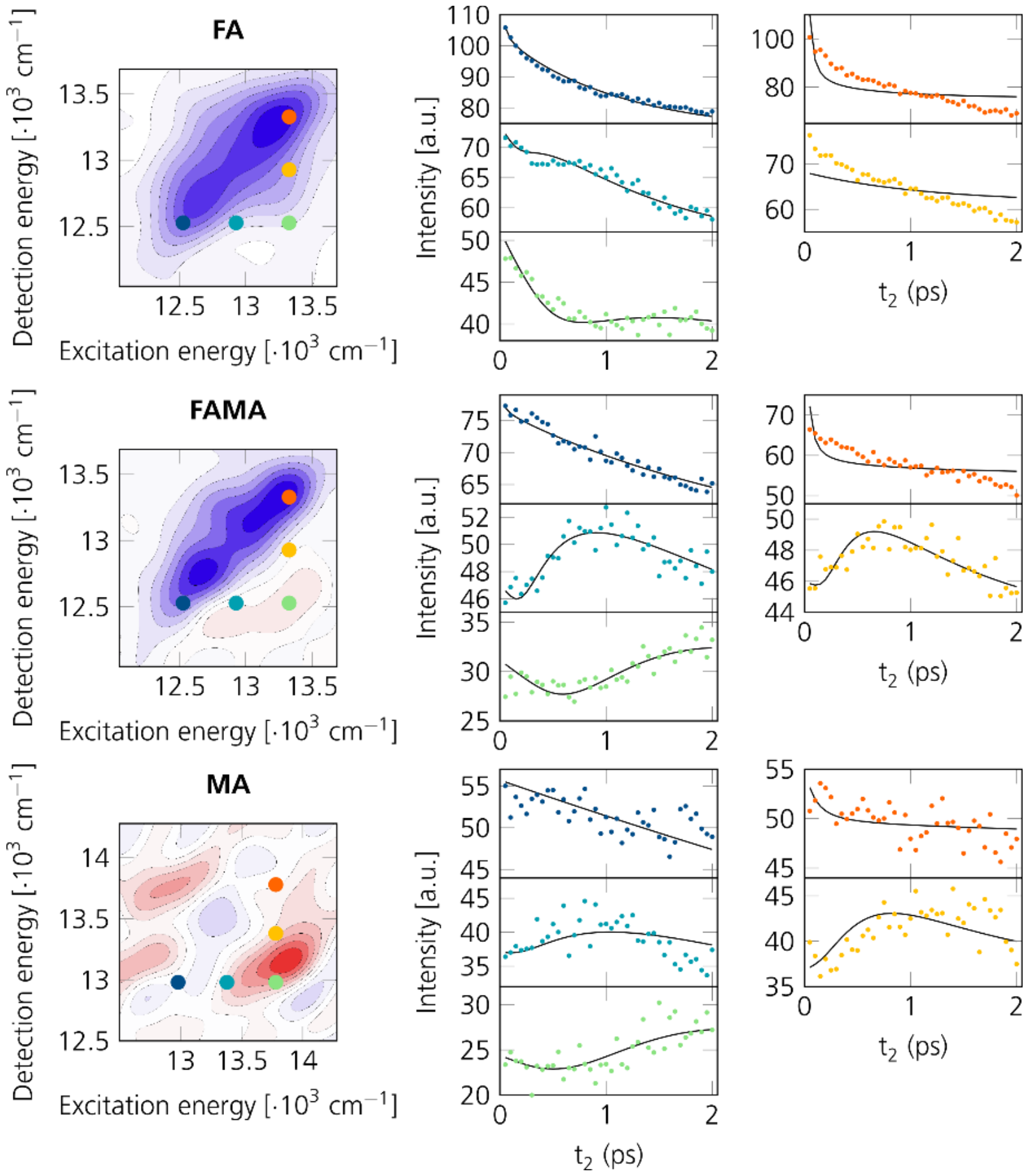

**Figure 4 | Tracking carrier dynamics via PC-2DES in different perovskite compositions**

Differential PC-2DES maps of the FA, FAMA and MA devices at $t_2$ = 2 ps are shown on the left. The coloured markers indicate the selected excitation–detection coordinates used for kinetic analysis. The corresponding population time traces are shown on the right, illustrating the carrier dynamics over the 50–2000 fs population time window. These traces are extracted from the original, non-subtracted PC-2DES data. Experimental data are shown as colored dots, while solid lines represent fits obtained from the kinetic model explained in the text.

The systematic increase of this delay with excitation excess energy provides direct experimental evidence for a cascade-like intraband cooling process. Rather than appearing instantaneously across the full spectral window, the low-energy photocurrent response builds up progressively as carriers redistribute through the conduction-band manifold. This behaviour is observed in all three compositions, indicating that sequential energy relaxation is a general feature of the working devices studied here.

The timescale of this buildup, however, depends on absorber composition. The $FAPbI_3$ device shows the most rapid development of the low-energy response following high-energy excitation, the mixed FAMA device exhibits intermediate behaviour, and the $MAPbI_3$ device shows the slowest rise. This trend shows that the rate at which excess carrier energy is dissipated differs systematically among the three absorber compositions. The observation is consistent with the role of A-site chemistry in tuning lattice structure, dynamic disorder and phonon-mediated relaxation pathways.[19,29]

The energy-resolved traces also show that the population dynamics contain several timescales: a prompt component related to the excitation, a delayed rise associated with intraband redistribution, and slower evolution of the signal amplitude at longer population times. The relative weight of these components varies with excitation and detection energy, indicating that the measured PC-2DES response contains both population transfer and additional spectral-response changes associated with the evolving carrier distribution.

These experimentally extracted kinetics establish the central result of the work: carrier cooling in complete perovskite solar cells is strongly composition dependent under operating conditions, with the relaxation sequence progressing from slowest in $MAPbI_3$ to fastest in $FAPbI_3$. We next use a kinetic model to test whether the observed energy-dependent dynamics can be described within a framework of phonon-mediated intraband scattering combined with population-dependent many-body effects.

### Kinetic interpretation of operando carrier dynamics

To obtain a compact physical interpretation of the measured kinetics, we employ an effective model rather than a unique microscopic description of the relaxation process. We represent the optically accessed conduction-band states as a discrete ladder of energy levels. Within this model, population transfer between these levels is described as phonon-mediated nonadiabatic transitions. Downward transitions correspond to phonon emission, whereas upward transitions are enabled by thermal phonon absorption.[30] The transition rates obey detailed balance, so that the carrier population relaxes toward a thermal distribution. In this picture, carriers excited with larger excess energy undergo a sequence of scattering events before contributing to the low-energy detection channels, naturally producing the delayed rise observed in the experimental traces.

The PC-2DES signal is also affected by population-dependent changes in the optical response that do not correspond directly to population transfer between ladder levels. In semiconductors, photoexcited carriers can induce additional dephasing and transient spectral shifts through many-body interactions. These effects modify the amplitude, linewidth and spectral position of features contributing to the photocurrent-detected response. We therefore include an additional spectral component to account for such many-body-induced spectral evolution. This term accounts for population-dependent spectral changes expected from excitation-induced dephasing and energy shifts as described by semiconductor Bloch equations.[31]

With this description, the model captures the main experimental trends across excitation-detection coordinates and absorber compositions. It captures the prompt response for near-resonant excitation, the delayed buildup of low-energy signal following higher-energy excitation, and the slower evolution of signal amplitude at longer population times. The model yields effective rise times for the population of the lowest-energy states of 1.7 ps in $MAPbI_3$, 1.3 ps in the mixed FAMA device and 1.0 ps in $FAPbI_3$. These values should be regarded as effective descriptors of the observed relaxation hierarchy rather than unique microscopic cooling times.

The model thereby provides a physically plausible description of the observed composition-dependent energy relaxation, while the present data do not uniquely identify the underlying microscopic scattering steps. Full details of the rate equations, phonon spectral density, many-body contribution, fitting procedure and robustness analysis are provided in SN 4-8.

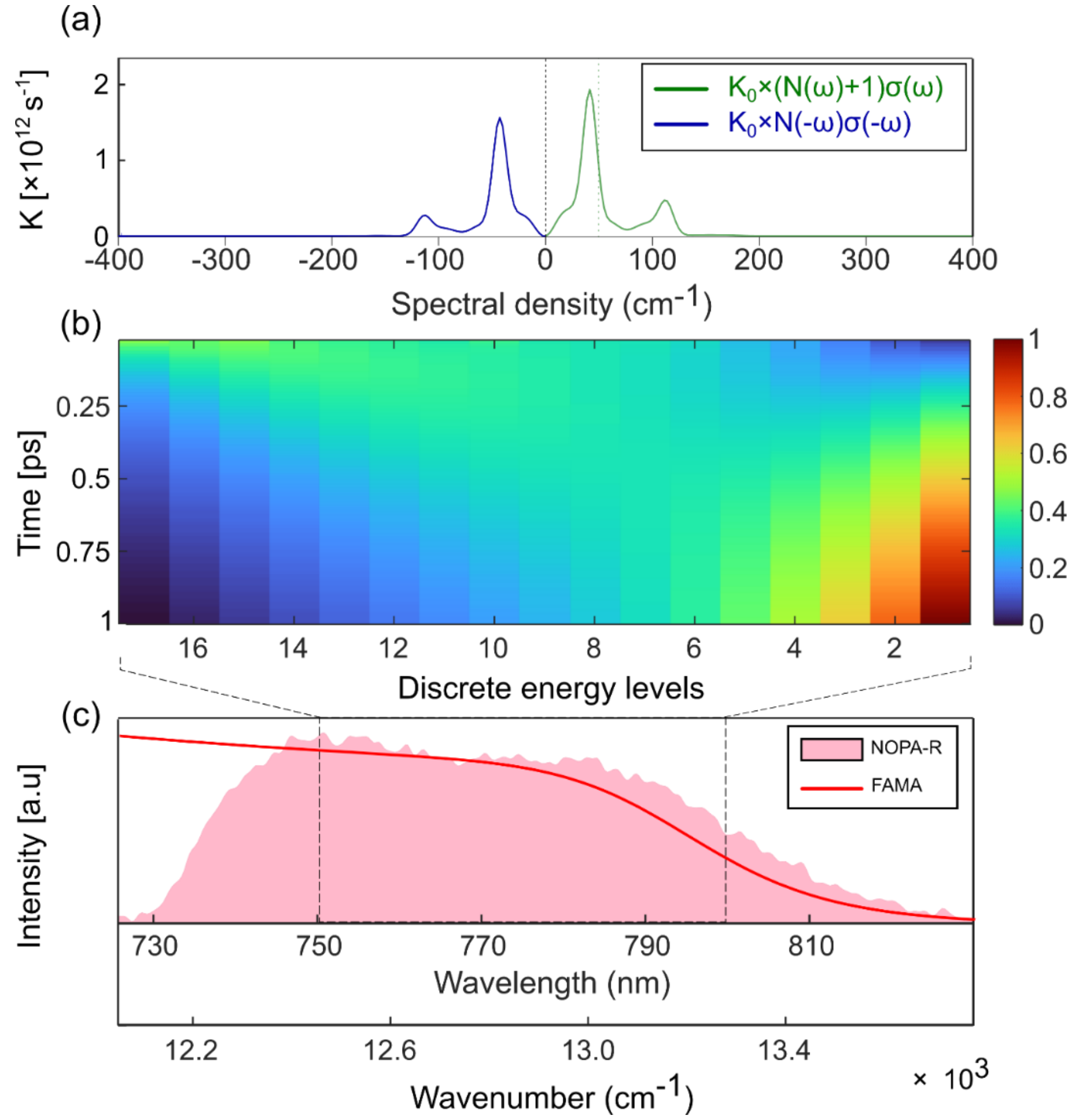

**Figure 5 | Kinetic interpretation of operando carrier relaxation.**

**a**, Simulated phonon-coupled spectral density-related energy gap dependent transfer rate in FAMA films, showing phonon emission (green) and reabsorption (blue) components. The spectral density is extracted based on resonance Raman and 2D THz electron-phonon coupling spectroscopy.[32–35] The downward and upward rates satisfy detailed balance.

**b**, Schematic of the effective conduction-band ladder and intraband carrier relaxation from the simulations. Carriers initially prepared at higher-energy levels relax toward lower-energy states through phonon-mediated scattering.

**c**, Excitation spectrum of the noncollinear optical parametric amplifier compared with the steady-state absorption spectrum of the FAMA device. The section of the spectrum that corresponds to the kinetic modelling is shown by thin dashed lines.

## Discussion

The present results show that carrier cooling in complete perovskite solar cells is not a universal material response but depends sensitively on absorber composition under operating conditions. Across all three devices, the PC-2DES maps evolve from an initially near-diagonal response toward below-diagonal spectral weight at later population times, providing an experimental signature of cascade-like redistribution from higher excitation energies toward lower-energy band-edge states.

The comparison between FA-, FAMA- and MA-based cells reveals a clear relaxation hierarchy: the low-energy response develops fastest in $FAPbI_3$, more slowly in the mixed FAMA device and slowest in $MAPbI_3$. This trend is captured by the reduced kinetic model, which describes the data using phonon-mediated redistribution through an effective conduction-band ladder together with a phenomenological many-body contribution. Because the model is an effective representation rather than a unique microscopic description, the fitted parameters should be interpreted as descriptors of the observed relaxation hierarchy.

The observed cooling hierarchy identifies an early-time contribution to device operation: the rate at which excess electronic energy is dissipated before carriers reach the band-edge states relevant for extraction. In this respect, the faster effective cooling observed in $FAPbI_3$ compared with $MAPbI_3$ is consistent with the prominent role of FA-rich absorbers in high-performance perovskite solar cells, although their device performance also depends strongly on phase stability, morphology and interface quality.[14,36] Rapid relaxation may shorten the residence time of carriers in highly excited states, thereby reducing access to hot-carrier-related interfacial loss channels and limiting non-equilibrium charge accumulation.[7,37] By contrast, the relatively small performance difference between FA and mixed FAMA devices is likely governed by additional materials and device factors, including film morphology, phase stability, defect density and contact quality, which are known to be strongly affected by processing and compositional stabilization strategies.[19,38] Thus, composition-dependent cooling represents one component of the performance-relevant energy-loss landscape, rather than a direct measure of solar-cell efficiency.

Although the band-edge states are largely derived from the inorganic Pb–halide framework, A-site substitution can modify the electronic structure indirectly[39] through lattice distortions, dielectric screening and coupling to Pb–halide lattice vibrations. In the present samples, the mixed FAMA absorber also contains a small Br fraction, so the observed relaxation hierarchy should not be attributed exclusively to A-site substitution. The observed trend therefore indicates that differences in absorber composition affect not only static optical properties, such as the bandgap, but also the ultrafast energy-flow processes that precede charge extraction.

These findings also suggest a possible link between early-time energy dissipation, device operation and stability. The transfer of excess electronic energy to the lattice may affect local heating, charge accumulation and internal electric fields, all of which are connected to non-radiative recombination, ion migration and degradation under illumination. The present measurements do not directly quantify degradation, but they identify composition-dependent relaxation pathways that may shape the physical conditions under which long-term operation occurs.

More broadly, this work establishes PC-2DES as a device-level probe of ultrafast processes in functional optoelectronic architectures. Because the nonlinear response is detected through photocurrent, the method is well suited to multilayer devices where conventional optical probes may be complicated by interference, scattering and weak differential transmission or reflection signals. The approach should therefore be applicable to other photovoltaic and optoelectronic systems in which early-time energy flow, charge separation and loss pathways need to be connected directly to device output.[40]

## Conclusion

In summary, PC-2DES reveals composition-dependent intraband carrier cooling in fully encapsulated perovskite solar cells. By resolving the evolution of photocurrent-detected spectral weight across excitation energy, detection energy and population time, we identify cascade-like relaxation from initially excited high-energy states toward lower-energy band-edge states. The cooling hierarchy is slowest in $MAPbI_3$, intermediate in the mixed FAMA device and fastest in $FAPbI_3$.

A reduced kinetic model incorporating phonon-mediated intraband scattering, supplemented by a phenomenological many-body contribution, captures the main energy-dependent trends and provides an effective parameterization of the observed relaxation dynamics. These results demonstrate that subtle changes in absorber composition can reshape the ultrafast pathways through which excess carrier energy is dissipated before charge extraction. They also establish action-detected multidimensional spectroscopy as a powerful platform for connecting ultrafast photophysics with device-relevant function in operating optoelectronic materials.

## Methods

### Device fabrication

This study explores perovskite solar cell devices incorporating three types of organic lead halide absorbers: $FAPbI_3$ (denoted as FA), $(FAPbI_3)_{0.992}(MAPbBr_3)_{0.008}$ (denoted as FAMA), and $MAPbI_3$ (denoted as MA). Fluorine-doped tin oxide (FTO)-coated glass substrates were patterned via laser scribing or chemically etched using a diluted HCl solution. The substrates were then sequentially cleaned with detergent, deionized (DI) water, acetone, and ethanol, followed by UV/ozone (UVO) treatment for 60 minutes prior to $SnO_2$ deposition. The $SnO_2$ electron transport layer was deposited by a chemical bath deposition (CBD) method. The precursor solution was prepared by dissolving 275 mg of $SnCl_2{\cdot}2H_2O$ (99.99%, Sigma-Aldrich), 1.25 g of urea (99.0%, Sigma-Aldrich), and 1.25 mL of concentrated HCl (37%, Sigma-Aldrich) in 100 mL of DI water under stirring for 30 minutes. FTO substrates were immersed in the solution and heated at 90 °C for 4 hours in an oven. After deposition, the films were rinsed with DI water and isopropanol (IPA) for 15 minutes each and subsequently annealed at 175 °C for 60 minutes.

Perovskite precursor solutions were prepared as follows:

- FA perovskite: 1.0128 g of pre-synthesized $FAPbI_3$ and 0.0386 g of MACl were dissolved in 1 mL of a mixed solvent of DMF (anhydrous, ≥99.8%, Sigma-Aldrich) and DMSO (anhydrous, ≥99.9%, Sigma-Aldrich) in an 8:1 volume ratio.
- FAMA perovskite: 1.0047 g of $FAPbI_3$, 0.0061 g of pre-synthesized $MAPbBr_3$, and 0.0386 g of MACl were dissolved in 1 mL of DMF/DMSO (8:1 v/v).
- MA perovskite: 0.2544 g of MAI and 0.7376 g of $PbI_2$ were dissolved in 1 mL of DMF/DMSO (8:1 v/v).

FA and FAMA solutions were spin-coated on $SnO_2$-coated FTO substrates in a two-step program at 1000 rpm for 5 s followed by 5000 rpm for 20 s. During the second step, 1 mL of diethyl ether was dripped onto the spinning substrate to induce adduct formation. The resulting films were annealed sequentially at 150 °C for 10 minutes and then at 100 °C for another 10 minutes. The MA solution was spin-coated at 4000 rpm for 20 s with 0.5 mL of diethyl ether dripped during spinning. These films were annealed at 65 °C for 1 minute, followed by 100 °C for 9 minutes. The hole transport layer was formed by spin-coating 30 µL of a spiro-OMeTAD solution (60 mg spiro-OMeTAD in 0.7 mL chlorobenzene, containing 25.5 µL 4-tert-butylpyridine and 15.5 µL of Li-TFSI solution [540 mg Li-TFSI in 1 mL acetonitrile]) at 4000 rpm for 20 s. All fabrication steps were performed under ambient atmosphere. Gold electrodes (70-100 nm) were thermally evaporated onto the spiro-OMeTAD layer under high vacuum ($<2\times10^{-6}$ Torr) at a deposition rate of approximately 0.5 Å/s. Device encapsulation was carried out inside a nitrogen-filled glovebox ($O_2/H_2O$ < 0.1 ppm) using a two-part epoxy (Gorilla Epoxy). Before encapsulation, the devices were left in the glovebox for at least 30 minutes to allow residual solvents to evaporate. A cover glass pre-coated with epoxy was gently placed over the device, and clamping pressure was applied to remove air bubbles. The encapsulated devices were then stored in the glovebox for a minimum of 12 hours to ensure complete curing of the epoxy.

**Device performance characterization**

*J-V* measurements were performed using a Keithley 2400 source meter under the simulated AM 1.5G one sun illumination (100 mW/cm$^2$) using a solar simulator (Oriel Sol 3A class AAA) equipped with 450 W Xenon lamp (Newport 6280NS). The light intensity was adjusted using a Si solar cell with KG-2 filter that was calibrated by National Renewable Energy Laboratory (currently National Laboratory of the Rockies). Absorption spectra were recorded by UV-vis spectrometer (Lambda 45, Perkin-Elmer) with the perovskite film coated on bare glass. Device performance parameters were verified to be consistent with high-performing perovskite solar cells for the respective compositions prior to ultrafast spectroscopy experiments. More details are available in the Supplementary Information.

**Photocurrent-detected two-dimensional electronic spectroscopy**

PC-2DES measurements were performed based on the methodology and apparatus detailed in our previous work.[41] A 1030 nm Yb:KGW laser (Pharos, Light Conversion ltd.) provided the fundamental beam, which subsequently pumped a home-built noncollinear optical parametric amplifier (NOPA), , operated in two configurations generating pulses of 100 nm bandwidth centered at 780 nm (NOPA-R) and 750 nm (NOPA-B). NOPA-R was used for the FA and FAMA devices and NOPA-B for the blue-shifted MA device. The pulses were temporally compressed down to 20 fs duration using a combination of chirped mirrors and a fused silica prism compressor for initial negative dispersion, followed by fine dispersion tuning (2nd, 3rd, and 4th order) using an acousto-optic programmable dispersive filter (Dazzler, Fastlite). The laser system operated at a 4 kHz repetition rate throughout all the experiments, while the streaming power of the Dazzler pulse shaper was set to 3%, giving energy per pulse of 1.75 nJ and distortion-free 2D maps. The set of 4 collinear phase-modulated pulses are then focused onto the sample by a spherical mirror down to a beam waist of 100 µm. The PC-2DES experiments involved scanning three time delays: the coherence time between the pump pulses ($t_1$) from 0 to 69 fs (step size 3 fs), the population time ($t_2$) between pump and probe pulse pairs from 0 to 2 ps (step size 50 fs), and the coherence time between the probe pulses ($t_3$) from 0 to 69 fs (step size 3 fs). The four-pulse sequence required for PC-2DES was generated using the Dazzler, which also implemented the phase modulation routine necessary for signal detection. Specifically, a 36-step phase pattern was applied, with the individual phases of the four pulses modulated at frequencies $f_1$=0 Hz, $f_2$=444.44 Hz, $f_3$=666.66 Hz, and $f_4$=999.99 Hz. Each 36-step phase pattern was repeated 80 times for each $t_1$, $t_2$, and $t_3$ point in the time-domain 3D-matrix to enhance the signal-to-noise ratio for subsequent Fourier analysis. Two-dimensional spectra were obtained by Fourier transformation with respect to the coherence times ($t_1$ and $t_3$).

For each absorber composition, two fully encapsulated working solar cells were selected for PC-2DES measurements. The photovoltaic performance of the selected devices was verified before the ultrafast measurements and was representative of the corresponding device batches. The two measured devices show very similar behaviour, see SN11. Comparison with modelling was carried out for one dataset for each composition.

**Data processing and analysis**

Different approaches to suppress incoherent population-mixing contributions in action-detected measurements have been developed recently.[26,42] In photovoltaic devices where charge separation and extraction are efficient, such contributions are expected to be reduced, although residual incoherent mixing cannot be excluded under the present excitation conditions (see SN3). Energy-resolved population dynamics were extracted by integrating the PC-2DES signal over selected excitation–detection energy pairs. All data processing and analysis were performed consistently across devices to enable direct comparison of relaxation dynamics between compositions.

**Kinetic modeling**

Kinetic interpretation of the experimental data was carried out using an effective multi-level model describing phonon-mediated intraband relaxation within the conduction band. Population transfer between discrete energy levels was governed by phonon emission and absorption processes which follow the spectral density of electron-phonon interaction and obey detailed balance. Population-dependent many-body effects were included phenomenologically to capture additional dynamical contributions to the photocurrent response. Numerical integration of the kinetic equations was used to simulate population dynamics and compare with experimental traces. Full details of the model, parameter selection, and optimization procedures are provided in the SN4-8.

**Data availability**

The data supporting the findings of this study are available from the corresponding author upon reasonable request.

**Code availability**

The analysis and modelling code used in this study are available from the corresponding author upon reasonable request.

**Acknowledgements**

E.A. and T.Pu. acknowledge financial support from the Swedish Energy Agency grant 50709-1, Olle Engkvist foundation grant 235-0422, the European Union's Horizon 2020 research and innovation program under the Marie Skłodowska-Curie grant agreement no. 945378 and no. 871124 Laserlab-Europe, the SNC Fellowship Program in Korea 2023, and the Royal Physiographic Society of Lund. L.B. and N.F.v.H. acknowledge support through the MCIN/AEI Projects PID2021-123814OB-I00, TED2021-129241B-I00, CEX2019-000910-S, Fundacio Privada Cellex, Fundacio Privada Mir-Puig, and the Generalitat de Catalunya through the CERCA program. E.A. and N.F.v.H. acknowledge financial support from PID2021-123814OB-I00 Light2Charge - Tracking Ultrafast Energy Transport on the Nanoscale. N.F.v.H. acknowledges support from ERC Advanced Grant 101054846-FastTrack. S.-H.L. and N.-G.P. acknowledge financial support from the National Research Foundation of Korea (NRF) through grants funded by the Korea government (MSIT and MOE) under contracts RS-2026-25501632 (NRL 2.0) and NRF-2021R1A3B1076723 (Research Leader Program).

**Author contributions**

E.A., Q.S. and T.P. drafted the manuscript with input from all authors. E.A. performed the PC-2DES measurements and analysis. Q.S. developed and performed the kinetic modelling. S.-H.L. fabricated and characterized the devices under the supervision of N.-G.P. L.B., D.Z. and N.F.v.H. contributed to PC-2DES methodology and interpretation. T.P. conceived and supervised the project. All authors discussed the results and commented on the manuscript.

**Competing interests**

The authors declare no competing interests.